\documentclass[smallextended]{svjour3}

\usepackage[T1]{fontenc}

\smartqed

\usepackage{graphicx}
\usepackage{mathptmx}      % use Times fonts if available on your TeX system
\usepackage{amssymb}

\journalname{Gen. Rel. Gravit.}

\begin{document}

%\preprint{AIP/123-QED}

\title{Energy Balance of a Bose Gas in a Curved Space-time}

\author{Tonatiuh Matos
	\and
	Ana Avilez
	\and
	Tula Bernal
	\and
	Pierre-Henri Chavanis
}

\institute{T. Matos \at Departamento de F\'{\i}sica,
 Centro de Investigaci\'on y de Estudios Avanzados del IPN,
 A. P. 14--740,  07000, Ciudad de M\'exico, M\'exico\\
 \emph{Part of the Instituto Avanzado de Cosmolog\'{\i}a
(IAC) collaboration http://www.iac.edu.mx/}\\
 \email{tmatos@fis.cinvestav.mx}
          \and
  A. Avilez \at Departamento de F\'{\i}sica,
 Centro de Investigaci\'on y de Estudios Avanzados del IPN,
 A. P. 14--740,  07000, Ciudad de M\'exico, M\'exico\\
 \emph{Present address: Facultad de Ciencias F\'sico Matem\'aticas,
 Benem\'erita Universidad Aut\'onoma de Puebla, Av. San Claudio y 18 Sur,
 Col. San Manuel, Edif. FM1-101B, Ciudad Universitaria, 72570, Puebla, M\'exico }\\
 \email{aavilez@fcfm.buap.mx}
		\and
 T. Bernal \at Departamento de F\'{\i}sica,
 Centro de Investigaci\'on y de Estudios Avanzados del IPN,
 A. P. 14--740,  07000, Ciudad de M\'exico, M\'exico\\
 \emph{Present address: Universidad Aut\'onoma Chapingo, km. 38.5 Carretera
       M\'exico-Texcoco, 56230, Texcoco, Estado de M\'exico, M\'exico}\\
 \email{ac13341@chapingo.mx}
		\and
 P.H. Chavanis \at
 Laboratoire de Physique Th\'eorique, Universit\'e Paul Sabatier,
 118 route de Narbonne 31062 Toulouse, France\\
 \email{chavanis@irsamc.ups-tlse.fr}
}

%\date{\today}
\date{Received: date / Accepted: date}
% The correct dates will be entered by the editor

\maketitle

\begin{abstract}
Classical solutions of the Klein-Gordon (KG) equation are used in astrophysics to model galactic halos 
of scalar field dark matter and compact objects such as
cores of neutron stars. These bound solutions are interpreted as
Bose-Einstein condensates whose particle number density is governed by the
Gross-Pitaevskii (GP) equation. It is well known that the Gross-Pitaevskii-Poisson
(GPP) system arises as the non-relativistic limit of the Klein-Gordon-Einstein (KGE) 
equations and, converselly, the KGE system may be interpreted as a generalization of the 
GPP equations in a curved space-time. 
In the present work, we consider a 3+1 ADM foliation of the space-time in order to
construct a general-relativistic version of the GP equation. Besides,
we derive a general energy balance equation for the boson gas in 
the hydrodynamic variables, where different
energy potentials are identified as kinetic, quantum, electromagnetic and 
gravitational. In addition, we find a correspondence between the energy potentials in the balance equation and actual components of the scalar energy-momentum tensor. We also study the Newtonian limit of the hydrodynamic formulation and the balance 
equation. As an illustrative case, we study the effects in the energy potentials due to a relativistic correction in the GP equation.
\end{abstract}

%\pacs{67.85.Hj, 67.85.Jk, 05.30.Jp, 04.40.Nr, 03.70.+k, 11.10.-z, 11.30.-j}

\section{Introduction}
\label{introduction}

Scalar fields are ubiquitous in modern physics, from the
inflaton responsible for the primordial acceleration of the Universe and the
Higgs scalar that gives mass to matter particles at fundamental
scales, up to huge astronomical and 
cosmological scales where they are used to model dark matter and dark energy. In most physical systems of interest, the dynamics of these scalars is well
described by the original Klein-Gordon-Maxwell (KGM) equations, which are
Lorentz and $U(1)$ invariant. It is well known that the KG modes can be 
interpreted as a set of independent bosonic particles living in a Minkowski
space-time. These bosons can be endowed with charge if a
complex field is considered,
whose electromagnetic interaction is mediated by a $U(1)$ gauge field. This is mathematically achieved by promoting the global $U(1)$ symmetry to a local one.
Self-interactions between bosons are encoded in a potential $V(\Phi)$.  

The first attempt to describe astronomical objects as
macroscopic bosonic states was made by 
\cite{Wheeler:1955} who aimed at constructing stable
particle-like solutions from
classical electromagnetic fields coupled to general relativity that he called
{\it geons} (an abbreviation for ``gravitational-electromagnetic entity'').
Later, \cite{Kaup:1968} and \cite{Ruffini:1969} introduced the
notion of boson stars that could be useful to model compact stars having certain
advantages over fluid  neutron stars models \cite{Faraoni:2010}. More recently, scalar fields as dark matter were suggested as a set of bosonic modes all
laying in the ground state making up a macroscopic wave function which
corresponds to a galactic halo 
(see e.g. \cite{Ruffini:1983,Spergel:1989,Sin:1994,Ji:1994,Lee:1995,Matos-Guzman:1998,Matos-Urena:2000,Hu:2000,Harko:2007,Chavanis:2011,Suarez-review:2014,Ostriker:2016},
among others). These studies suggest that
these objects are gigantic 
Bose-Einstein condensates (BECs). In a general context, these configurations
are classical solutions of the Klein-Gordon-Einstein (KGE)
system and 
their evolution and stability have been widely studied in the last decades. 
  
 Due to the Bose-Einstein statistics, all individual particles
lay in a common quantum state and therefore the macroscopic wave 
 function scales as its occupation number. Consequently, the
corresponding mean-field non-linear Schr\"odinger equation, called the
Gross-Pitaevskii (GP) equation, encodes the evolution of the number density of the condensate. It has been
shown in a host of works that the Schr\"odinger-Poisson (SP) and
Gross-Pitaevskii-Poisson (GPP) equations arise as the
non-relativistic limit of the Klein-Gordon-Einstein (KGE) equations (see
\cite{Chavanis-Matos:2016} for a review). Therefore, it is common to
interpret the KGE solutions as densities of number of particles of
bosonic systems laying in a curved space-time background. Usually, such
non-linear system is solved numerically and  the existence of bounded stable
soliton-like solutions, dubbed as boson stars, has been demonstrated.
 
 The stability theory of these configurations can be understood from the
elliptical nature of the equations. A simple and intuitive idea is the
following: because these solutions are self-gravitating and dispersive by nature, the balance between these competing features determines the stability
of these objects \cite{Seidel-Suen:1990,Seidel-Suen:1991,Seidel-Suen:1994}. In
these works, it has been shown from numerical full-relativity calculations 
that the critical maximum mass of a scalar configuration depends on the
inverse of the mass of the boson. This implies that for heavy
bosons, bound solutions cannot be of astrophysical size. Besides, they found that the 
stability of an initial configuration is determined by its initial total mass and the
initial central value of the scalar field $\phi_0^i$. For small initial central values of 
the density,  stable bound solutions along the whole range 
of masses are possible. After passing a critical value of the $\phi_0^i$, a small branch of 
solutions collapse into a black hole and finally, when the 
total rest mass of the bosons exceeds the gravitational binding energy for larger values of 
$\phi_0^i$, dispersive solutions cannot be hold together and hence they fade away.

In the equilibrium regime, these boson stars solutions have been used to model
dark matter halos of galaxies for a long time. From the last three
decades the implications of these sort of models of dark matter have been
analyzed in a systematic manner in different regimes (see e.g. \cite{Ruffini:1983,Spergel:1989,Sin:1994,Ji:1994,Lee:1995,Matos-Guzman:1998,Matos-Urena:2000,Hu:2000,Harko:2007,Chavanis:2011,Suarez-review:2014,Ostriker:2016}, among others). A 
particularly interesting feature of 
this model is that it naturally predicts the non-existence of small substructures in 
contrast to the cold dark matter (CDM) model and therefore this model
overcomes some potential issues of CDM such as the missing satellite problem, for example.

Another interesting application of this model is the formation of black holes from the 
collapse of unstable boson stars. This mechanism has been used 
to explain the formation of black holes of different kinds in the Universe, such as 
supermassive black holes \cite{Herdeiro:2014,Escorihuela:2017,CruzGuzman:2011}. Despite
the fact that the no-hair theorems condemn stable scalar field configurations
to exist around black holes, it has been shown that within this scenario the existence of 
scalar quasi-resonant solutions that decay very slowly is possible, and they can be used to 
model actual galactic dark matter halos surrounding supermassive black holes 
\cite{Barranco:2011,Barranco:2012,Escorihuela:2017,Avilez:2017}. This idea provides a 
consistent mechanism of formation of supermassive black holes even
at very early stages in the history of the Universe \cite{Avilez:2017}.
It is also important to study the scalar fields 
in the strong field regime, specially after the discovery of gravitational waves which opens 
a new quest about strong gravity. In the context of boson
stars, the previous quest leads us to face an outstanding  and
unsolved problem in science, that is, finding exact general solutions of the
KGE equations. Although such a task has been unreachable so far, approximations have
always allowed us to turn around impossible tasks in physics, 
specially if one aims to make applications in astrophysics. An approximate and commonly used 
approach is to assume a static scenario at which the
gravitational dynamics have reached an equilibrium point and the problem of
solving the complex KGE system is turned over. The static solutions in this limit
suffice to correctly describe several systems.

One of the most intriguing problems in general relativity is to
describe the dynamics of matter in strongly relativistic regimes such as gases,
fluids and fields near compact objects like neutron stars or black holes. The
standard hydrodynamical theory does not allow us to identify the different
energy contributions because the standard 
laws of thermodynamics are not applicable in fully relativistic environments. 
Essentially, the problem stems from the fact that the space-time metric 
describes both the geometrical structure and the dynamical aspects of the system
\cite{Wald:1984}.

 In this article, as a first step, we address this problem for a charged boson gas at zero temperature, described by the KGM Madelung transformation of the scalar field. In this new field variables, the KG equation for the complex scalar field is transformed 
into a couple of hydrodynamic equations governing the dynamics of the boson gas.
In this representation, it is easy to identify the different contributors to the total energy of the system through a balance
equation that is an antecedent of the
first law of thermodynamics in a curved space-time for a quantum gas made of bosons. In addition, we find some relations between the energy potentials arising in the balance equation and the actual expected values of the energy and momentum for the system, quantified by the components of the energy-momentum tensor of the scalar field defined into the 3+1 foliation. 
 
 Although all this KG machinery has been the basis of great achievements 
as, e.g. the inflaton, the $\pi-$mesons, the Higgs
boson, axions, etc., further
developments of this framework are needed to model phenomena beyond the Minkowskian threshold. In this work, we are interested in solutions of the KG
equation as models of objects at large scales like compact stars and dark matter halos.

  In the non-relativistic limit, the KG equation endowed with a self-interaction reduces to the GP
equation. Since it is based on Newtonian gravity, the GP equation cannot model compact objects. However, it is possible to construct a more suitable
framework by extending the GP equation to curved space-times,
which is an important goal of this work. This article is
organized as follows.
In section~\ref{sec:field-eqs} we present the field equations describing our system of bosons.  In section~\ref{sec:generalGP} we 
generalize the GP equation, which provides a model for a macroscopic system of charged bosons laying in a curved space-time. In section~\ref{sec:hydrodynamics}, we
derive the hydrodynamic representation for the complex scalar field equations by a field redefinition using Madelung variables. In section~\ref{sec:Euler} we derive 
the generalized Euler and continuity equations from the complex Klein-Gordon equation. In section~\ref{sec:balance-eq} we derive a conservation equation that can be
interpreted as the energy balance of the different components involved. 
In section~\ref{sec:energy-tensor} we compute some ADM invariants formally interpreted as the energy and momentum defined along a generic $3+1$ foliation. As usual, these quantities are written in terms of the components of the energy-momentum tensor and some other geometrical entities. By performing the Madelung field-redefinition, these quantities are written in terms of the energy potentials. Such relation provide valuable information helpful about the physical interpretation of the hydrodynamic energy potentials.
In section~\ref{sec:Newtonian} we derive the Newtonian limit of the balance equation and its energy
components. In section~\ref{sec:1st-correction} we present our simple case of study of a scalar field in flat space-time with a first-order relativistic correction. Finally, in section~\ref{conclusions} we present our conclusions.

\section{Field Equations}
\label{sec:field-eqs}

  In what follows we model the boson gas
as a set of excitations of a self-interacting, 
charged, complex scalar field which is minimally coupled to a gauge vector field
mediating 
the electromagnetic interaction. Gravity is interpreted as a
geometrical phenomenon, i.e. the 
surrounding space-time of a massive body acquires
curvature as described by General 
Relativity. We shall not deal explicitly with the Einstein
field equations but we consider an arbitrary space-time geometry. We aim to
extend the KGM equations, with local $U(1)$
symmetry,  by using coordinates for a $4$-dimensional manifold playing the role
of the physical curved space-time whose geometry is encoded by a metric $g$.
From hereafter we use the units $c=hbar=\epsilon_0=\mu_0=1$, for $c$
the speed of light, $hbar$ the reduced Planck constant, $\epsilon_0$ and
$\mu_0$ the permittivity and permeability of free space, respectively. We
define the electromagnetic d'Alembert operator as $\Box_\mathrm{E} \equiv \left(\nabla^{\mu} + i e A^{\mu} \right) \left( \nabla_{\mu} + i e A_{\mu} \right)$, where $e$ is the charge unit
and $A_{\mu}$ is the $U(1)$ gauge vector field corresponding to the Maxwell 4-potential, such
that the KGM equations are given by 
\begin{eqnarray}
	&&\Box_\mathrm{E} \Phi - \frac{\mathrm{d}V}{\mathrm{d}\Phi^*} = 0 ,
    \label{eq:KG} \\
 &&\nabla_\nu F^{\nu\mu} =J^{E\,\mu} ,
\label{eq:Maxwell}
\end{eqnarray}
for the complex scalar field $\Phi(t,\mathbf{x})$ and its complex conjugate
$\Phi^*(t,\mathbf{x})$. The Faraday tensor is given by
\begin{equation}
	F_{\mu\nu} = \nabla_\mu A_\nu - \nabla_\nu A_\mu \, ,
\end{equation}
and the conserved 4-current is defined as
\begin{equation}
	J^E_\mu \equiv i\frac{e}{2 m^2} \left[\Phi
\left(\nabla_\mu - i e A_\mu \right) 
    \Phi^*  - \Phi^* \left( \nabla_\mu + i e A_\mu \right) \Phi \right] .
\label{eq:Jmu}
\end{equation}
We introduce scalar self-interactions by using the ``$\Phi^4$'' self-interacting potential
\begin{equation}
	V(\Phi) =  m^2 |\Phi|^2 + \frac{\lambda}{2} |\Phi|^4 ,
\label{eq:V}
\end{equation}
describing a system of bosonic 
excitations that condenses into a single macroscopic ground state.

  We use the $3+1$ ADM foliation of the metric
\cite{ADM:2008,Alcubierre:2008} so that the
line-element reads
\begin{equation}
	\mathrm{d}s^2 = - N^2 \mathrm{d}t^2 + \gamma_{ij} \left(\mathrm{d}x^i + N^i \, \mathrm{d}t \right) \left(
	\mathrm{d}x^j + N^j \, \mathrm{d}t \right) ,
\label{eq:ds}
\end{equation}
where $N$ is the lapse function which measures the proper time of the observers
traveling along the world line, $N^i$ is the shift vector that measures the displacement
of the observers between the spatial slices and $\gamma_{ij}$ is the spatial
metric.

\section{Generalized Gross-Pitaevskii Equation}
\label{sec:generalGP}

Here we show that the KG equation with the hat potential~(\ref{eq:V}) can be
transformed into a relativistic  GP equation. The non-relativistic GP equation,
which is a non-linear version of Schr\"odinger's
equation, has been of great interest in quantum
and statistical mechanics since it accounts for correlations between
quantum
particles.  An important application of this framework is the study of
superfluidity and phase transitions. \cite{Bogolyubov:1947} first tried to
model superfluidity as an imperfect BEC due to the weak repulsion between the
bosons. It is worth noting that, in contrast to the standard GP equation for
neutral particles in a spatially flat space-time,
charged bosons in a curved space-time find interesting
applications in the context of dark matter, boson
stars and neutron stars with superfluid cores. 

  An important mathematical feature of the KG equation is that, in some circumstances, it 
admits non-dispersive solutions, as those relevant in scattering processes. The Derrick 
theorem states that non-regular, static, non-topological localized scalar field solutions
are stable in a spatially flat space \cite{Derrick:1964}. This constraint is avoided by
adopting a harmonic decomposition for the complex scalar field,
\begin{equation}
	\Phi(t,\mathbf{x}) = \Psi(t,\mathbf{x}) \exp(-i\omega_0 t) \, ,
\end{equation}
where $\omega_0$ is a constant that can be either the mass or the frequency of massless 
particles. Although the field is non-static, the space-time remains static and
thus the
KG equation admits soliton-like solutions \cite{Rosen:1966,Derrick:1964}. By plugging
such field-redefinition into the KG equation~(\ref{eq:KG}) we
obtain
\begin{eqnarray}
	&&i \nabla^0 \Psi - \frac{1}{2\omega_0} \Box_E \Psi 
    + \frac{1}{2\omega_0} \left(  m^2 + \lambda n \right) \Psi 
     \nonumber\\
&&	+\quad \frac{1}{2} \left( - \frac{\omega_0}{N^2}
	- 2 e A^0 + i \, \Box \, t \right) \Psi = 0 ,
\label{eq:GP}
\end{eqnarray}
where $ n(t,\mathbf{x})\equiv|\Phi|^2 = |\Psi|^2$ is defined as the scalar field
density 
and $\Box \, t = \nabla^\mu \nabla_\mu t$.\footnote{In order to define a
slicing
condition to calculate the lapse function $N$, a common approach uses harmonic 
coordinates, defined by asking for the wave operator to vanish: $\Box \, x^\alpha = 0$. 
In particular, for $\Box \, t = 0$ the \emph{harmonic slicing condition} on the lapse 
function is obtained. Here we do not assume any specific foliation condition.}
Equation~\ref{eq:GP} is the KG equation in terms of $\Psi$ and we name it the 
\textit{generalized GP} equation in curved space-times for the
potential~(\ref{eq:V}).

\section{Hydrodynamic Representation}
\label{sec:hydrodynamics}
  
  The hydrodynamic approach for the Schr\"odinger equation was introduced by Madelung 
\cite{Madelung:1927} (see also \cite{Bohm:1952a,Bohm:1952b}), 
who showed that it is equivalent to Euler's equations for an irrotational fluid with an
additional quantum potential. 
The boson gas that we study can be interpreted as a real fluid described by the quantum Euler equations.
This hydrodynamic representation has been
widely used in the literature for BEC dark matter \cite{Harko:2007,Chavanis:2011,Suarez:2015,Bettoni:2014}, including
electromagnetic interactions \cite{Chavanis-Matos:2016}, and for
relativistic BEC stars \cite{Chavanis-Harko:2011,Chavanis:2014}. In this section, we 
derive the hydrodynamic representation for the generalized GP equation~(\ref{eq:GP}). We
carry out the following Madelung transformation:
\begin{eqnarray}
	\Phi(t,\mathbf{x}) = \sqrt{n} \exp(i\theta) =
\sqrt{n}\exp \left[i(S - \omega_0 t)\right],
\label{eq:definicion}
\end{eqnarray}
where the amplitude $\Psi$ is decomposed into a density $n(t,\mathbf{x})$ and a phase
$S(t,\mathbf{x})$ encoding the geometry and evolution of the front-wave 
solution. In this way, the KG/GP equation splits into its imaginary and real
parts, respectively:
\begin{equation}
	\nabla_{\mu} \sqrt{n} \left( 2 \nabla^{\mu} \theta  + e A^{\mu} \right)
    + e \nabla_{\mu} \left( A^{\mu} \sqrt{n} \right) + \sqrt{n} \, \Box \theta = 0 ,
\label{GP:imag}
\end{equation}
\begin{equation}
	\Box \sqrt{n} - \sqrt{n} \left[ \nabla_{\mu} \theta \left( \nabla^{\mu}
\theta  + 2 e A^{\mu} \right) + e^2 A^2 +m^2 + \lambda n \right] = 0 ,
	\label{GP:real}
\end{equation}
where $A^2 = A^\mu A_\mu$. After applying the Madelung transformation, the 
current~(\ref{eq:Jmu}) turns into
\begin{equation}
 	J^E{}_{\mu}= \frac{n e}{m^2} \left( \nabla_{\mu}
\theta + e A_{\mu} \right).
 \label{GP:J}
 \end{equation}
  Interestingly, a relativistic quantum particle in a flat space-time with
electromagnetic field has a mechanical momentum
$m \mathbf{u} = \mathbf{p} - e 
\mathbf{A}$,  with $\mathbf{u}$ the 4-velocity, $\mathbf{p}$ the canonical momentum
and $\mathbf{A}$ the magnetic
vector potential. By writing its wave function in hydrodynamic variables, it results that 
$\mathbf{p}= \nabla S$.
In a similar 
way for our boson gas, the electromagnetic $4$-momentum corresponds to the sum
of 
individual mechanical momenta, namely $ 
J^E{}_{\mu} = ( e/ m) n u_\mu$. In terms of 
$J^E{}_{\mu}$,
equations~(\ref{GP:imag}) and~(\ref{GP:real}) read
\begin{eqnarray}
	\nabla^\mu J^E{}_\mu = 0 ,
\label{GP:imag:J}\\
	J^E{}_{\mu}J^E{}^\mu  + \frac{n^2e^2}{{ m}^4}
\left(  m^2 + \lambda n
	- \frac{\Box \sqrt{n}}{\sqrt{n}} \right) = 0 .
\label{GP:real:J}
\end{eqnarray}
Then, by interpreting the KG equation as a general-relativistic GP equation,
through the Madelung transformation it splits into the continuity~(\ref{GP:imag:J}) and
quantum Hamilton-Jacobi~(\ref{GP:real:J}) equations above. Such quantum version 
of the Hamilton-Jacobi equation differs from the classical one only by the last term on
the left-hand side of equation~(\ref{GP:real:J}), that corresponds to
the de Broglie relativistic quantum potential \cite{DeBroglie:1927}.

  Now, let us take the continuity equation~(\ref{GP:imag:J}) and notice that
\begin{equation}
	\int_{\mathbb V} \nabla_{\mu} J^{E\mu} \, \mathrm{d}V =
    \int_{\mathbb V} \nabla_0 J^{E0}  \, \mathrm{d}V +
    	\int_{\mathbb S} k_j J^{Ej}  \, \mathrm{d}S = 0 ,
\label{eq:int}
\end{equation}
where $\mathbb{S}$ is an arbitrary surface enclosing the volume $\mathbb{V}$ containing
the system and $k^j$ is the vector orthogonal to $\mathbb{S}$. We assume that far away
from the sources $J^E{}_j$ is negligible, so we are free to
choose a volume $\mathbb{V}$
such that the surface integral of equation~(\ref{eq:int}) vanishes. Thus, the
quantity $Q =
\int_\mathbb{V} J^{E0} \, \mathrm{d}V$ is a conserved charge,
\begin{eqnarray}
	\frac{\mathrm{d}Q}{\mathrm{d}t} = \int_\mathbb{V}\nabla_0 J^{E0} \,
\mathrm{d}V = 0 .
\end{eqnarray}
In conclusion, the continuity equation~(\ref{GP:imag:J})
expresses the
conservation of the charge of the scalar field.

\section{Continuity and Euler Equations}
\label{sec:Euler}
 
 Let us define the ``velocity'' $v_\mu$ of an individual particle as
\begin{equation}
	 m v_\mu \equiv \nabla_\mu S +  e A_\mu .
\label{eq:vel}
\end{equation}
It is important to mention that $v_\mu$ is not a 4-vector strictly speaking.
We also stress that we are working in a frame in which the
contribution of the rest-mass energy has been subtracted from the
4-velocity.
  In terms of $v_\mu$, the continuity and quantum
Hamilton-Jacobi equations~(\ref{GP:imag:J}) and~(\ref{GP:real:J}) become
\begin{equation}
	\nabla^\mu (nv_\mu) - \frac{\omega_0}{ m}  \left( \nabla^0 n + n \, \Box \,
    t \right) = 0 ,
\label{eq:n1}
\end{equation}
\begin{equation}
	v_\mu v^\mu - \frac{2 \omega_0}{m} v^0 - \frac{\omega_0^2}{ m^2 N^2} +
   1 + \frac{\lambda}{ m^2} n - \frac{1}{ m^2}\frac{\Box\sqrt{n}} {\sqrt{n}} = 0
.
\label{eq:n2}
\end{equation}
Equation~(\ref{eq:n1}) governs the evolution of the density of the boson gas
whilst~(\ref{eq:n2}) governs the evolution of its phase.

After applying the covariant derivative $\nabla_\alpha$ to
equation~(\ref{eq:n2}), using the Leibniz rule to the
first two terms and using Maxwell's equations~(\ref{eq:Maxwell}), we
obtain the Euler equation 
\begin{eqnarray}
	&&v_\mu \nabla^\mu v_\alpha - \frac{\omega_0}{ m} \nabla^0 v_\alpha
	- \frac{\omega_0^2}{2 m^2} \nabla_\alpha \left(\frac{1}{N^2}\right) 
    \nonumber\\
	&+& \frac{\lambda}{2 m^2} \nabla_\alpha n - \frac{1}{2 m^2} 
    \nabla_\alpha
	\left( \frac{\Box \sqrt{n}}{\sqrt{n}} \right) \label{eq:Hydro2} \\
	&+& \frac{ e}{ m} \left[ v_\mu(\nabla_\alpha A^\mu - \nabla^\mu A_\alpha)
	- \frac{\omega_0}{ m}(\nabla_\alpha A^0-\nabla^0 A_\alpha) \right] = 0 . 
    \nonumber
\end{eqnarray}
At this point, we can identify different analogues to physical quantities. In first
place, we spot the covariant definition of the Lorentz force in a curved
space-time, given by 
\begin{eqnarray}
	F^E{}_\alpha \equiv -\frac{ e}{ m}\left(v_\mu F_\alpha{}^\mu-\frac{\omega_0}
    { m} F_\alpha{}^0\right).
\label{eq:Lorentz}
\end{eqnarray}
In second place, we identify the gravitational ``force'' which, in spite of the
geometrical nature of gravity assumed here,  is an actual
measurement of the curvature
associated with the gravitational strength quantified by the time-time component of the
metric
\begin{equation}
     F^G{}_\alpha \equiv - 2\nabla_\alpha U^G; \qquad  U^G \equiv
     -\frac{\omega_0^2}{4N^2 m^2} ,
\end{equation}
where $U^{G}$ is the gravitational ``potential''
contribution due to the
metric time-component related to the gravitational strength.
In third place, the quantum force is given by
\begin{equation}
	F^Q{}_\alpha\equiv-\nabla_\alpha U^Q; \qquad U^Q \equiv
- \frac{1}{2 m^2}\frac{\Box\sqrt{n}}{\sqrt{n}} ,
\label{eq:Bohm-force}    
\end{equation}
where $U^Q$ is the quantum potential. Finally, a measure of the temporal and spatial 
variations of the density of the scalar field is characterized
by 
\begin{equation}
  F^n{}_\alpha \equiv - \nabla_\alpha h; \qquad h\equiv\frac{\lambda n}{2  m^2} ,
\label{eq:N-force}
\end{equation}
where $h$ is the enthalpy. Notice that $F^n{}_\alpha$ is only present when the self-interactions are turned-on
($\lambda \neq 0$).

For future purposes, we introduce the definition for the pressure $p\equiv
{\lambda}n^2/4m^2$, which satisfies the Gibbs-Duhem relation
$\mathrm{d}h = \mathrm{d}p/n$, and the internal 
energy $U^n={\lambda}n/4m^2$, which satisfies the local law
$\mathrm{d}U^n=-p \ \mathrm{d}(1/n)$.

  In summary, if all the previous quantities~(\ref{eq:Lorentz}-\ref{eq:N-force}) 
are plugged into Euler's equation~(\ref{eq:Hydro2}), we obtain
\begin{equation}
	-\frac{\omega_0}{ m} \nabla^0 v_\alpha + v_\mu \nabla^\mu v_\alpha =
	F^E{}_\alpha + F^G{}_\alpha + F^Q{}_\alpha + F^n{}_\alpha \, .
\label{eq:forces}
\end{equation}
Equations~(\ref{eq:n1}), (\ref{eq:n2}) and~(\ref{eq:forces})
are dynamically equivalent to the KGM 
equations. However, written in terms of the $n$ and $v_\mu$ variables,
they give rise to a
different physical interpretation. They may be viewed as the
generalized continuity, Hamilton-Jacobi and Euler
hydrodynamic equations.

\section{Balance Equation}
\label{sec:balance-eq}
  
  We now derive the different energy contributions for the system of 
charged bosons and the total energy balance equation. Notice that contracting
equation~(\ref{eq:forces}) with $nv_{\alpha}$ and using
the Leibniz rule we get
\begin{equation}
	\nabla^\mu (v_\mu n K) - \frac{\omega_0}{ m} \nabla^0(nK) 
	= n v^\mu (F^E{}_\mu + F^G{}_\mu + F^Q{}_\mu + F^n{}_\mu) ,
\label{eq:Energia}
\end{equation}
where $K \equiv (1/2) v_\alpha v^\alpha$ is defined as the kinetic energy per unit mass.

  In order to express the electromagnetic contribution in terms of
the symmetric energy-momentum tensor, it is convenient to introduce the
current of charge
\begin{equation}
J^{E\mu}=\frac{ e}{ m}n\left (v^{\mu}-\frac{\omega_0}{
m}\nabla^{\mu} t\right ).
\label{cur}
\end{equation}
With this current, the Lorentz force~(\ref{eq:Lorentz}) takes the form
\begin{equation}
F^E{}_{\alpha}=-\frac{1}{n}J^{E\mu}F_{\mu\alpha}.
\end{equation}
Using the covariant Maxwell equations~(\ref{eq:Maxwell}), after some tensor
algebra and using the Jacobi identity of the Riemann tensor, we obtain
\begin{equation}
F^{E\alpha}=\frac{1}{n}\nabla_{\beta}\Theta^{\alpha\beta},
\label{nice1}
\end{equation}
where the symmetric electromagnetic energy-momentum tensor is defined as
\begin{equation}
	\Theta^{\alpha\beta} \equiv  g^{\alpha\mu}F_{\mu\nu}F^{\nu\beta}
    + \frac{1}{4} g^{\alpha\beta} F^{\mu\nu} F_{\mu\nu}  \, .
\end{equation}
Its components $\{\Theta^{00},\Theta^{0i}\} = \{ U^{E},
P^{E}_i\}$ are
the generalized electromagnetic energy density and the Poynting
vector, respectively.

  On the other hand, using equation~(\ref{cur}) the
first term on the r.h.s.~of equation~(\ref{eq:Energia}) reads
\begin{equation}
n v^{\alpha}F^E{}_{\alpha}=\left
(\frac{ m}{ e}J^{E\,\alpha}+\frac{\omega_0}{
m}n\nabla^{\alpha} t\right )F^E{}_{\alpha}.
\label{nice2}
\end{equation}
The antisymmetry of the Faraday tensor implies that
$J^{E\alpha}F^E{}_{\alpha}\propto J^{E\alpha}J^{E\mu}F_{\mu\alpha}=0$, which
leads us to obtain from equations~(\ref{nice1})
and~(\ref{nice2}) the following:
\begin{equation}
	nv^\alpha F^E{}_\alpha=  -\frac{\omega_0}{
m}nF^{E0}= -\frac{\omega_0}{
m}\nabla_\alpha\Theta^{\alpha 0} \, .
\label{eq:Lorentz2}
\end{equation}
Since equation~(\ref{eq:Lorentz2}) is a gauge
invariant expression, the tensor structure of our results below will not be affected by the
gauge fixing procedure used on the electromagnetic sector. 

  Now, using the continuity equation~(\ref{eq:n1}) and the Leibniz rule, it is
easy to get the following relation for an arbitrary field $U$:
\begin{eqnarray}\nonumber
	-\frac{\omega_0}{m}\nabla^0(n U)&+&\nabla^{\mu}(v_\mu n 
    U) -n v_\mu\nabla^\mu U\\
    +\frac{\omega_0}{m}
    n \nabla^0 U &+& \frac{\omega_0}{m}
    n U\Box t =0 ,
\label{eqs:conservation}
\end{eqnarray}
which can be applied to the internal energy $U^n$, the quantum potential $U^Q$
and the gravitational contribution $U^G$.

 A further simplification is made by noting that
\begin{equation}
	n \nabla^0 U^Q = - \frac{1}{4  m^2} \nabla_\mu \left[ 
	n \nabla^0 \left(\nabla^\mu \ln n\right) \right] .
\end{equation}
This relation, along with equation~(\ref{eqs:conservation}), leads to 
\begin{eqnarray}
	-\frac{\omega_0}{ m} \nabla^0(nU^Q) &+& \nabla^\mu \left( n v_\mu U^Q + 
    J^Q{}_\mu \right) \nonumber \\
    - n v_\mu \nabla^\mu U^Q&+& \frac{\omega_0}{ m}n U^Q \Box t = 0 ,
\label{eq:UQ}
\end{eqnarray}
where we have defined the \textit{quantum flux} as
\begin{equation}
	J^Q{}_\mu\equiv -\frac{\omega_0}{4 m^3} n \nabla_\mu ( \nabla^0 \ln n ).
\end{equation}
If we sum equations~(\ref{eq:Energia}) and (\ref{eqs:conservation})
for $U^n$ and
$U^G$, and equation~(\ref{eq:UQ}), we obtain
\begin{eqnarray}\nonumber
	&-& \frac{\omega_0}{ m} \nabla^0(n \mathcal{U}^s) + \nabla^\mu(n v_\mu 
    \mathcal{U}^s) + \frac{\omega_0}{ m} \nabla^iP^E{}_i + \frac{\omega_0}{ m} \nabla^0 U^E   
     \\
     \nonumber
    &+& \nabla^\mu \left( J^Q{}_\mu + p v_\mu \right)
+\frac{\omega_0}{m}n\nabla^0 U^G +
nv^{\mu}\nabla_{\mu}U^G\\ &+&\frac{\omega_0}{
m}n\left (U^G+U^Q\right ) \ \Box t = 0 \, ,
\label{eq:EnergiaT}
\end{eqnarray}
where we have introduced the energy density of the scalar field, $\mathcal{U}^s
\equiv K + U^G + U^Q + U^n$. Equation~(\ref{eq:EnergiaT}) is
the total energy balance equation.
The total flux associated
with the energy density of the scalar field $\mathcal{U}^s$ involves the energy
flux $n v_\mu 
    \mathcal{U}^s$, the quantum flux $J^Q{}_\mu$, the pressure flux $p
v_\mu$, and a contribution due to the 
gravitational interaction $U^G$. The flux associated with the electromagnetic
energy density $U^E$ is the Poynting
vector $P^E{}_i$.

\section{The Energy-Momentum Tensor}
\label{sec:energy-tensor}

  By construction, in general relativity, the energy and momentum densities of matter are usually defined as the time-time and time-space of the expected value of the energy momentum tensor laying at the r.h.s.~of Einstein's equations. The goal of this section is to find out a relation between such components of the energy-momentum tensor and the different so-called ``energy'' contributions appearing in the balance equation for the boson gas studied here.  

  The energy-momentum tensor for a Bose gas reads
\begin{eqnarray}\nonumber
T_{\mu\nu}&=&T^\Phi_{\mu\nu}+T^A_{\mu\nu}\nonumber\\
&=&\frac{1}{2}\left[\left(\Phi_{,\mu}+iA_\mu\right)\left(\Phi_{,\nu}^*-iA_\nu\right)\right. \nonumber\\
&+&\left(\Phi_{,\mu}^*-iA_\mu\right)\left(\Phi_{,\nu}+iA_\nu\right)\nonumber\\
&-&\left.g_{\mu\nu}\left(\left(\Phi_{,\sigma}+iA_\sigma\right)\left(\Phi^{*,
\sigma}-iA^\sigma\right) +V\right)\right] .
\label{eq:EnergyTensor1}
\end{eqnarray}
As we shall see shortly, for the sake of clearness, we split it
in two terms:
\begin{eqnarray}
	T^v_{\mu\nu} &=& n\left[ \left(\ln\sqrt{n} \right)_{,\mu} \left(\ln\sqrt{n}\right)_{,\nu}-\frac{1}{2}g_{\mu\nu} \left(\ln\sqrt{n}\right)_{,\sigma} \left(\ln\sqrt{n}\right)^{,\sigma}\right.\nonumber\\
&+& \left(\theta_{,\mu}+eA_\mu\right) \left(\theta_{,\nu}+eA_\nu\right)\nonumber\\
&-&\left.\frac{1}{2}g_{\mu\nu} \left(\left(\theta_{,\sigma}+eA_\sigma\right)\left(\theta^{,\sigma}+eA^\sigma\right)+\frac{V}{n}\right)\right],
\label{eq:EnergyTensorP1}
\end{eqnarray}
and 
\begin{eqnarray}
	T^\theta_{\mu\nu} &=& e\sqrt{n} \bigg[ \left( \left(\ln\sqrt{n}\right)_{,\mu}A_{\nu}+\left(\ln\sqrt{n}\right)_{,\nu}A_{\mu}\right.\nonumber\\
&-& \left.g_{\mu\nu}(\ln\sqrt{n})_{,\sigma}A^{\sigma}\Big)\sin\theta\right.\\
&+& \left(\theta_{,\mu}A_{\nu}+\theta_{,\nu}A_{\mu}-g_{\mu\nu}\theta_{,\sigma}A^{\sigma}\right)\left(\cos\theta-\sqrt{n}\right)\bigg].\nonumber
\label{eq:EnergyTensorA1}
\end{eqnarray}
Rewritten in terms of the velocities $v_\mu$, equation~(\ref{eq:EnergyTensorP1}) transforms into
\begin{eqnarray}
	T^v_{\mu\nu} &=& n \left[ \left(\ln\sqrt{n}\right)_{,\mu} \left(\ln\sqrt{n}\right)_{,\nu} - \frac{1}{2}g_{\mu\nu}\left(\ln\sqrt{n}\right)_{,\sigma}\left(\ln\sqrt{n}\right)^{,\sigma}\right.\nonumber\\
&+&m^2v_{\mu}v_{\nu}-\frac{1}{2}g_{\mu\nu}\left(m^2v_{\sigma}v^{\sigma}+\frac{V}{n}\right)\\
&-&\omega_0\left(m v_\mu\delta_\nu^0+m v_\nu\delta_\mu^0+\omega_0 \delta_\mu^0\delta_\nu^0-g_{\mu\nu} \left(m v^0-\frac{\omega_0}{2N^2}\right)\right)\bigg].\nonumber
\label{eq:EnergyTensorP}
\end{eqnarray}
At this point, the motivation to decompose
$T_{\mu\nu}$ into $T_{\mu\nu}^v$ and $T_{\mu\nu}^\theta$ becomes clear since it
is not possible to write $T_{\mu\nu}^\theta$ in terms of the velocity:
\begin{eqnarray}\nonumber
	T^\theta_{\mu\nu} &=& e\sqrt{n} \bigg[ \left( \left(\ln\sqrt{n}\right)_{,\mu}A_{\nu}+\left(\ln\sqrt{n}\right)_{,\nu}A_{\mu}\right.\\
&-&g_{\mu\nu}\left(\ln\sqrt{n}\right)_{,\sigma}A^{\sigma}\Big)\sin\theta\nonumber\\
&+&\bigg(m\left(v_{\mu}A_{\nu}+v_{\nu}A_{\mu}-g_{\mu\nu}v_{\sigma}A^{\sigma}\right)\\
&-&\omega_0 \left(\delta_\mu^0 A_\nu+\delta_\nu^0A_\mu-g_{\mu\nu}A^0\right)\nonumber\\
&-&\left.\left.2e \left(A_\mu A_\nu+\frac{1}{2}g_{\mu\nu}A_\sigma A^\sigma\right)\right) \left(\cos\theta-\sqrt{n}\right)\right].\nonumber
\label{eq:EnergyTensorA}
\end{eqnarray}

  Let us introduce a covariant 4-vector $n_\mu$ laying in a 3-dimensional hyper-surface given by $n_\mu=(-N,0,0,0)$ and a contravariant 4-vector $n^\mu=\frac{1}{N}\left(1,N^i\right)$, such that $n_\sigma n^\sigma=-1$. Let us recall that $N$ is the lapse function of the $3+1$ foliation (see equation~(\ref{eq:ds})). Additionally, let us introduce $h_i^\mu=\delta_i^\mu+n_i n^\mu$ which is a  projector tensor. In our case, it corresponds to $h_i^\mu=\delta_i^\mu$. Thus, in terms of these quantities, the scalar field density is defined as
\begin{eqnarray}
  \rho^v&=&n_{\mu}n_{\nu}T^{\Phi\mu\nu}\\
  &=& n\left[N^2 \left(\nabla^0\ln\sqrt{n} \right)^2 + \frac{1}{2} \left(\ln\sqrt{n}\right)_{,\sigma}\left(\ln\sqrt{n}\right)^{,\sigma} + \frac{V}{2n}\right.\nonumber\\
  &+& \left. m^2N^2 \left(v^0\right)^2 + \frac{1}{2} m^2 v_{\sigma}v^{\sigma} + \omega_0 \left(m v^0-\frac{\omega_0}{2N^2}\right) \right], \nonumber
\label{eq:SFDensity}
\end{eqnarray}
and 
\begin{eqnarray}
  \rho^\theta&=&n_{\mu}n_{\nu}T^{A\mu\nu}\nonumber\\
  &=& e \left[N^2 \left(\nabla^0\sqrt{n}\right)A^0  + \left(\sqrt{n}\right)_{,k}A^{k} \right]\sin\theta \nonumber\\
  &+& e \sqrt{n}\left[m \left(N^2v^0A^0+v_{k}A^{k}\right) - \omega_0\left(1-\frac{2}{N}\right)A^0\right.\nonumber\\
  &-& e \left(N^2\left(A^0\right)^2+A_kA^k\right) \left(\cos\theta-\sqrt{n}\right)  \bigg].
\label{eq:SFDensity0}
\end{eqnarray}
In the same way, we can obtain the fluxes 
\begin{eqnarray}
  \mathrm{J}^v_i&=&n^{\mu}h^\nu_i T^{\Phi}_{\nu\mu}\\
&=& n\left[-N\left(\nabla^0\ln\sqrt{n}\right)\left(\ln\sqrt{n}\right)_{,i}-m^2Nv^0v_i-\frac{\omega_0m v_i}{N}\right],\nonumber\\
S^v_{ij} &=& h^\nu_i h^\mu_j  T^\Phi_{\mu\nu}\\
&=& n\left[\left(\ln\sqrt{n}\right)_{,i}\left(\ln\sqrt{n}\right)_{,j}-\frac{1}{2}\gamma_{ij}\left(\ln\sqrt{n}\right)_{,\sigma}\left(\ln\sqrt{n}\right)^{,\sigma}\right.\nonumber\\
&-&\frac{1}{2}\gamma_{ij}\left(m^2+\frac{\lambda}{2}n\right)\nonumber\\
&+&\left.m^2v_iv_j-\frac{1}{2}\gamma_{ij}v_{\sigma}v^{\sigma}+\gamma_{ij}\omega_0\left(m v^0-\frac{\omega_0}{2N^2}\right)\right],\nonumber
\label{eq:SFluxiP}
\end{eqnarray}
and
\begin{eqnarray}
\mathrm{J}^\theta_i &=& n^{\mu}h^\nu_i T^{A}_{\nu\mu}\\
&=& - eN \bigg[\left(\nabla^0\sqrt{n}A_{i}+\left(\sqrt{n}\right)_{,i}A^0\right) \sin\theta\nonumber\\
&+& \sqrt{n}\left.\left(m\left( v^0A_i+v_iA^0\right)+\frac{\omega_0}{N^2}A_i+\frac{2e}{N}A^0A_i\right)\right.\nonumber\\
&\times& \left(\cos\theta-\sqrt{n}\right)\bigg],\nonumber\\
S^\theta_{ij} &=& h^\mu_i h^\nu_j T^A_{\mu\nu}\\
&=& \bigg[\left( \left(\sqrt{n}\right)_{,i}A_{j} + \left(\sqrt{n}\right)_{,j}A_{i} - \gamma_{ij}\left(\sqrt{n}\right)_{,\sigma}A^{\sigma}\right)\sin\theta\nonumber\\
&+& \sqrt{n}e\left.\bigg( m \left(v_iA_j+v_j A_i-\gamma_{ij}v_{\sigma}v^{\sigma}\right)\right.\nonumber\\
&+& \left.\left.\gamma_{ij}\omega_0A^0
-2e\left(A_i A_j-\frac{1}{2}\gamma_{ij}A_\sigma A^\sigma\right) \right)\left(\cos\theta-\sqrt{n}\right)\right].\nonumber
\label{eq:SFluxiA}
\end{eqnarray}
The main goal of this section is to figure out the relation of the different energy contributions in the balance equation~(\ref{eq:EnergiaT}) to the actual densities of energy and momentum derived from the energy-momentum tensor. The motivation of doing so is to give physical meaning to such quantities.    
For convenience, we write the expressions above in terms of the energy potentials in the balance equation~(\ref{eq:EnergiaT}). In order to do so, notice that the Bernoulli equation~(\ref{eq:n2}) leads to 
\begin{eqnarray}
\frac{\omega_0}{m}v^0=K+2U^n+2U^G+U^Q+\frac{1}{2}.
\label{eq:Bernoulli2}
\end{eqnarray}
In addition, notice that the following relations are satisfied
\begin{equation}
\left(\ln\sqrt{n}\right)_{,\sigma} \left(\ln\sqrt{n}\right)^{,\sigma}
= -N^2\left(\nabla^0\ln\sqrt{n}\right)^2+\frac{n_{,k}n^{,k}}{4n^2},
\end{equation}
\begin{equation}
2m^2U^Q = -\frac{\Box n}{2n}+\left(\ln\sqrt{n}\right)_{,\sigma}\left(\ln\sqrt{n}\right)^{,\sigma} ,
\end{equation}
which can be used to write the actual energy density as
\begin{eqnarray}
\rho^v &=& nm^2\left(1+2\tilde{U}\right) -\frac{n}{4}\Box\ln n \nonumber\\
&-&\left[U^n-2U^Q+\frac{1}{4U^G}\left(K+\frac{1}{2}\right)^2\right] ,
\label{eq:SFDensity2}
\end{eqnarray}
where we have defined $\tilde{U} \equiv K+2U^n+2U^G+U^Q$. We can follow a similar procedure in order to relate the fluxes to the energy potentials:
\begin{equation}
S^v_{ij}=\frac{n_{,i}n_{,j}}{4n}-\frac{1}{2}\gamma_{ij}\Box n
+m^2n\left[v_iv_j-\gamma_{ij}\left(U^n+4U^G\right)\right].
\label{eq:SFluxi2}
\end{equation}
In order to layout an interpretation for the relations above notice that~(\ref{eq:SFDensity2}), which corresponds to the energy density of the system according to the ADM formalism, is related to the rest mass of the gas and the total energy. It is worth to mention that the influence of the rest of the components could be important. Equation~(\ref{eq:SFDensity2}) shows that the contribution of each class of energy is non-trivial and that the energy-momentum tensor contains a number of components whose relation with physical conserved quantities is not explicit a priori.

\section{Newtonian Limit}
\label{sec:Newtonian}
  
  In the longitudinal Newtonian gauge in flat space the interval is given by
\begin{equation}
	\mathrm{d}s^2 = -(1-2\varphi) \mathrm{d}t^2 + \delta_{ij} (1+2\varphi)\, 			
    \mathrm{d}x^i\mathrm{d}x^j ,
\end{equation}
 which implies that the lapse function is $N^2=1-2\varphi$, and $\gamma_{ij}=(1+2\varphi)\delta_{ij}$.
Within the non-relativistic limit, equation~(\ref{eq:GP})
becomes the traditional GP equation~\cite{Pitaevskii:2003} and the
hydrodynamic equations~(\ref{eq:n1}), (\ref{eq:n2}) and~(\ref{eq:forces}) 
reduce to
\begin{eqnarray}
	&\partial_t n + \nabla\cdot (n {\bf v})=0 , \label{eq:momentum1} \\
	&\partial_t S + \frac{1}{2 m}\left(\nabla S - e{\bf A}\right)^2 = - m \left( U^Q + h
	+ \varphi + \frac{e}{m} \varphi_E \right), \nonumber \\
\label{eq:momentum3} \\
	&\partial_t {\bf v}+\left({\bf v}\cdot \nabla\right){\bf v} = - \nabla U^Q
    - \frac{1}{n}\nabla p-\nabla\varphi+\frac{e}{m}\left({\bf E}+{\bf v}\times {\bf B}\right),
    \nonumber \\
	\label{eq:momentum2}
\end{eqnarray}
where $U^Q=-(1/2m^2)\Delta\sqrt{n}/\sqrt{n}$
is the Madelung classical quantum 
potential~\cite{Madelung:1927} and $({\bf E},{\bf B})=(-\partial_t{\bf
A}-\nabla\varphi_E,\nabla\times {\bf A})$ is the
electromagnetic field.

Taking the scalar product of equation~(\ref{eq:momentum2}) with ${\bf v}$ and
using the
continuity equation~(\ref{eq:momentum1}), we obtain
\begin{eqnarray}
	\partial_t \left (n K\right )+\nabla \left (n K {\bf
	v}\right )=&-&n{\bf v}\cdot \nabla U^Q-{\bf v}\cdot \nabla p\nonumber\\
	&-&n{\bf v}\cdot \nabla \varphi + {\bf J}^E\cdot {\bf E},
\label{en}
\end{eqnarray}
where $K=v^2/2$ is the density of kinetic energy and we have
introduced the current of charge ${\bf J}^E=(e/m)n{\bf v}$. From the Maxwell
equations $\nabla\times {\bf E}=-\partial_t{\bf B}$ and
$\nabla\times {\bf B}={\bf J}^E+\partial_t{\bf E}$, using the identity
$\nabla\cdot ({\bf E}\times {\bf B})={\bf B}\cdot (\nabla\times
{\bf E})-{\bf E}\cdot (\nabla\times {\bf B})$, one can show that ${\bf J}^E\cdot {\bf
E}=-\partial_tU^E-\nabla\cdot {\bf P}^E$, where $U^E=(E^2+B^2)/2$ is the 
electromagnetic energy and
${\bf P}^E={\bf E}\times {\bf B}$ is the Poynting vector. From the continuity 
equation~(\ref{eq:momentum1}), we get
\begin{equation}
	\partial_t(nU)+\nabla(nU{\bf v})=n\partial_tU+n{\bf v}\cdot \nabla U.
\label{id}
\end{equation}
Applying equation~(\ref{id}) to $U^Q$, $U^n$ and
$U^G=\varphi/2$, using the identity 
\begin{equation}
	n\partial_t U^Q=-\nabla\cdot {\bf J}^Q; \qquad {\bf J}^Q
    =\frac{1}{4m^2} n \ \partial_t\nabla \ln n,
\end{equation}
and introducing the energy density of the  scalar field ${\cal
U}^s = K+U^Q+U^n+U^G$,
we obtain the local energy conservation equation
\begin{eqnarray}
	\partial_t(n\mathcal{U}^s)&+&\nabla\cdot \left
	(n{\cal U}^s{\bf v}\right )+\partial_tU^E+\nabla\cdot
	{\bf P}^E+\nabla\cdot (p{\bf v}) \nonumber\\
	&+& \nabla\cdot {\bf J}^Q + \frac{1}{2}n {\bf
	v}\cdot\nabla\varphi
	-\frac{1}{2}n\partial_t\varphi=0 .
\label{eq:NewtonBalance}
\end{eqnarray}
Using the Poisson equation, $\nabla^2\varphi = 4\pi Gm^2n$, and the continuity
equation~(\ref{eq:momentum1}), one can easily show that~(\ref{eq:NewtonBalance}) implies the
global conservation of the total (scalar field $+$
electromagnetic) energy contribution, $E_{\rm tot}=\int \left(n{\cal U}^s+U^E\right)\, \mathrm{d}
{\bf x}$, i.e. $\dot E_{\rm tot}=0$.
  This equation relates different 
contributions of the energy of a set of bosons laying within a gravitational potential well 
deforming the Minkowski space-time. This approximation suffices to describe a
host of astrophysical situations.

\section{Relativistic first-order correction in time to the Newtonian approach}
\label{sec:1st-correction}

As shown in previous sections, the Newtonian limit of the KGM system in its hydrodynamical representation is obtained by choosing a particular foliation of the space-time, meanwhile terms of higher order in powers of $v/c$ are neglected.
In this section, we present an application of our generalized GP equation and
its corresponding hydrodynamic representation in order to illustrate how a
first-order correction in powers of $v/c$ gives rise to new features of the
different components in the balance equation. By now, we ignore the
electromagnetic fields in order to focus 
on the effects due to general relativity. The modified behavior
of the solutions presented here, carries up new physics
inherited by the general relativistic nature of the underlying
KG system from which this limit is obtained.

In this section, we present two simple examples in the weak gravity and diluted density limits. Firstly, the simplest  planar-wave-like solution and secondly, we consider the simplest spherically symmetric solution assuming a uniform sphere of bosonic gas. Both cases are simple academic examples that pretend to illustrate the behavior of the potentials appearing in the balance equation.

\subsection{Weak gravity limit}

Let us start by assuming that the gravitational potential well generated by the
boson gas is weak enough, i.e. $\varphi(t,r)\ll 1$, such that the space-time is
Minkowskian with a good approximation. In this case the constraint Einstein
equations reduce to the Poisson equation given by
\begin{equation}
  \nabla^2\varphi=4\pi G \Phi\Phi^*.
\label{eq:Einstein_r}
\end{equation}

We start with a simple ansatz, by considering the case
of a planar wave solution for $\Phi$, i.e. 
\begin{equation}\label{eq:planaransatz}
\Phi=\Phi_0\exp\left(ik_\mu x^\mu\right) = \Phi_0\exp\left[i(\mathbf{k}\cdot
\mathbf{x}-\omega t)\right].
\end{equation}
The KG equation~(\ref{eq:KG}) reduces to a dispersion relation
\begin{eqnarray}
	(1-2\varphi)\omega^2-(1+2\varphi)k^2
	-(m^2+\lambda n)=0,\label{eq:Planar}
\label{eq:PlanarW1}
\end{eqnarray}
and the corresponding Poisson equation~(\ref{eq:Einstein_r}) turns into
\begin{eqnarray}
	\nabla^2\varphi=4\pi G \left(m^2+\frac{\lambda}{2}n\right)n .
\end{eqnarray}
By now, let us assume the simplest scenario in which the scalar configuration is made of a homogeneous gas configuration with $n=\Phi_0^2=$constant.
In principle, equation~(\ref{eq:Planar}) provides a dispersion relation for the
planar wave inside the self-generated gravitational well. Notice that by
ignoring self-gravity effects the dispersion relation for a free plane-wave is
recovered:  $k^2=\omega^2-(m^2+\lambda n)$. At this point, it is worth to notice
that a massless-boson-like behavior, $k=\omega$, might be obtained if the
density of the boson gas satisfies $n =-m^2/\lambda$. Such situation is only
possible for $\lambda<0$, that is, when bosons 
in the gas repel each other. The previous has physical
meaning since self-gravity of the bosons is compensated by a repulsive contribution from pressure. Under such conditions, the
gravitational potential for the bosonic system is given by
\begin{equation}
	\varphi=\frac{1}{3} \pi Gm^2 n r^2 - \frac{C_1}{r}+C_2 ,
\label{eq:varphir}
\end{equation}
where $r$ is the radial coordinate. In this case the velocity in the
hydrodynamic formulation is given by  $m v_\mu=\omega_0\delta_\mu^0$, since
$S=k_\mu x^\mu+\omega_0 t$. 
 After making the previous assumptions, the kinetic, gravitational and internal energy potentials are given by
\begin{eqnarray}
	K &=& \omega_0^2(1-2\varphi),\\
	U^G &=& \frac{\omega_0^2}{4m^2}(1-2\varphi)=\frac{K}{4m^2},\\
	U^n &=& - \frac{1}{4}.
\label{eq:energyCont}
\end{eqnarray} 

The behavior of the scalar solution is linked to the
evolution of the potentials above. Since $\varphi$ is a periodic function of
time and space, the kinetic and gravitational potentials
also are. As a consequence, these contributions should compensate each other
into the balance equation. On the other hand, $U^n$ is
constant. This is expected because we are dealing with a flat space-time.

\subsection{Diluted density limit}

Other simple case that might be relevant for different applications in astrophysics is to consider spherically symmetric solutions of the KG equation in a weak gravity regime. Consider that the density of bosonic particles is diluted, so that $n=n(r,t) \ll 1$. Under such assumptions, the KG equation~(\ref{eq:KG}), using hydrodynamic variables, reduces to
\begin{equation}
  \frac{\Box\sqrt{n}}{\sqrt{n}} = \left[m^2-\left(\mu-\omega_0\right)^2\right] + i\left(\mu-\omega_0\right)\frac{n_t} {n} ,
\label{eq:KG_r}
\end{equation}
where the phase has been harmonically decomposed as $S=\mu t$, and $n_t$, $n_r$ %
denote derivatives with respect to time and radius, respectively. For consistency, we set $\mu=\omega_0$ in equation~(\ref{eq:KG_r}) to have a real solution of the density. This last equation is solved by the following radial profile:
\begin{equation}
  n(t,r)=\frac{1}{r^2}\left[\sqrt{C_1^2+C_2^2}+C_1\sin(2kr)+C_2\cos(2kr)\right]T(t).
\label{eq:n_r1}
\end{equation}
 We choose the integration constants in~(\ref{eq:n_r1}) in order to get a regular non-singular solution given by
\begin{equation}
  n=n_0\frac{\sin^2(kr)}{k^2r^2}\cos^2(\omega\,t),
\label{eq:n_r}
\end{equation}
with the dispersion relation $\omega\equiv\sqrt{k^2+m^2}$.
Notice that the previous solution for the density is a modified version of the
Newtonian solution given by $n=n_0 \sin^2(kr) /(kr)^2$, where a periodic
time-dependence has arisen due to 
first-order relativistic corrections.
At this point, the gravitational potential 
generated by such a solution for the scalar field can be computed by plugging
the previous solution for $n$ into (\ref{eq:Einstein_r}), resulting in
\begin{equation}
 \varphi = \frac{\varphi_0}{k^2}\left[\ln(2kr)+\frac{1}{2k}\sin(2kr)
- {\rm Ci}(2kr)\right]\cos^2(\omega\,t) ,
\label{eq:L_r}
\end{equation}
where $\varphi_0\equiv 8\pi G m^2 n_0^2$ and ${\rm Ci}(x)$
is the cosine integral function given by
\begin{equation}
 {\rm Ci}(x)=\gamma+\ln(x)+\int_0^x
\frac{\cos(y)-1}{y}\mathrm{d}y,
\end{equation}
being $\gamma=0.5772$.

It is worth to point out that the gravitational 
potential has a harmonic evolution in time in contrast to the Newtonian case
which is static.
However, the quantum potential of the bosonic system defined by~(\ref{eq:Bohm-force})
turns out to be constant, $U^Q=-2$, which implies that the quantum force is
equal to zero. Nevertheless, the quantum flux $J^Q_\mu$ evolves non-trivially
and its components (assuming spherical symmetry) are given by
\begin{eqnarray}
J^Q_r&=&\frac{(1+\varphi)\omega_0}{4m^3 n}\left(\frac{n_tn_r}{n}- n_{tr}+n_t\varphi_r\right),\\
J^Q_t&=&\frac{(1+\varphi)\omega_0}{4m^3 n}\left(\frac{n_tn_t}{n}-n_{tt}\right).
\label{eq:QFlux_r}
\end{eqnarray}

\begin{figure}
\centering
  \includegraphics[width=0.8\columnwidth]{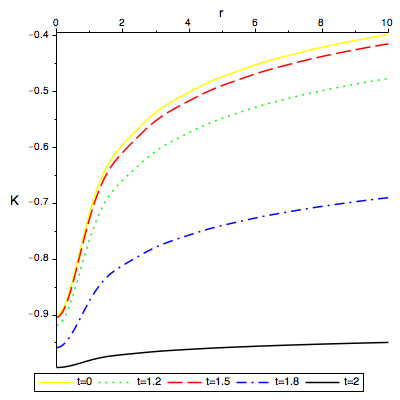}
  \caption{The behavior of the Kinetic potential $K$ of the system. The potential is bigger for outer side of the system. The values of the parameters  used in the plot are $\mu=1,\,
\omega_0=0.1,\,k=2,\,m=1,\,n_0=1$.}
 \label{fig:Kin_r}
\end{figure}

\begin{figure}
\centering
  \includegraphics[width=0.8\columnwidth]{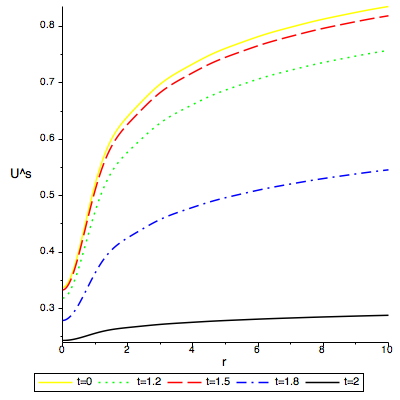}
  \caption{The behavior of the total energy contribution $\mathcal{U}^s$
of the
system. The flux is bigger in the outside
side of the system. Values of the parameters used in this plot are the same as in Fig.~\ref{fig:Kin_r}.}
 \label{fig:Us_r}
\end{figure}

\begin{figure}
\centering
  \includegraphics[width=0.8\columnwidth]{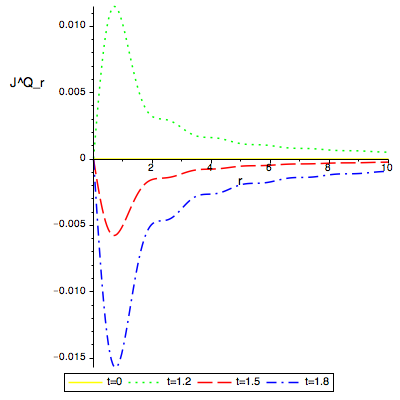}
  \caption{The behavior of the radial component of the quantum flux.}
 \label{fig:Flux_r}
\end{figure}

\begin{figure}
\centering
  \includegraphics[width=0.8\columnwidth]{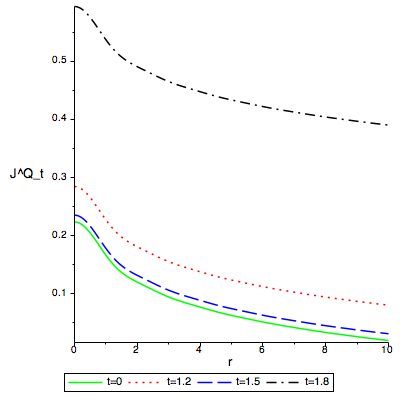}
  \caption{The behavior of the $t$-component of the quantum flux.
The flux is bigger at the center of the
system. The values of the plot are the same as in Fig.~\ref{fig:Kin_r}}
\label{fig:Flux_t}
\end{figure}

Snapshots of the time evolution and of the spatial structure
of the components of the quantum flux are illustrated in Figs.~\ref{fig:Flux_r}
and \ref{fig:Flux_t}.
On the other side, the Bernoulli equation, which is equivalent to the KG equation~(\ref{eq:KG_r}) after being expressed in hydrodynamic variables, is useful to compute the kinetic potential given by
\begin{equation}
2K=v_\mu v^\mu=-\frac{\omega_0^2}{m^2}\left(1-\varphi\right),
\label{eq:Kinetic_r}
\end{equation}
which clearly evolves non-trivially in time, see in Figs.~\ref{fig:Kin_r}. and has almost the same dependence on the
gravitational potential than the kinetic potential for the planar wave. For comparison   Fig.
\ref{fig:Us_r} shows snapshots of the total energy contribution 
$\mathcal{U}^s$ of the system.

See in Figs.~\ref{fig:Us_r}, \ref{fig:Flux_r} and \ref{fig:Flux_t} the
corresponding 
behaviors of the quantum flux and the total energy contribution $\mathcal{U}^s$ of the system.

\section{Conclusions}
\label{conclusions}
  
  In this  article, we have derived a generalized hydrodynamic formulation of a system of 
charged bosonic excitations laying in a curved space-time, which is governed by a 
general relativistic set of continuity, Hamilton-Jacobi, and
Euler equations written up in 
Madelung variables, equivalent to the Klein-Gordon-Maxwell equations.
By performing a $3+1$ foliation of the space-time we are able to handle curved geometries within this framework. We have shown
that it is possible to split
the total energy contribution of the hydrodynamical system associated to the boson gas into different contributions or potentials, specifically the
kinetic, quantum and
electromagnetic parts, and a term due to the gravitational field strength arisen from the curvature of space-time. The main
result of this article is the energy balance equation for the
boson gas which plays the role of an hydrodynamical first law of thermodynamics for the system in the general
relativistic regime.
In addition, in order to relate the  potentials involved in the balance equation, we compute some physical conserved quantities defined along the $3+1$ foliation such as the total energy and the projected momenta
which are  written in terms of the energy-momentum tensor of the scalar and pure geometric entities associated with the foliation. In this way, we establish a mapping between the potentials and actual physical observables. 
It is worth
remarking that this result is 
 general and has not been
derived before and it is an important result for models involving canonical scalar fields in astrophysics, such as models of dark matter and neutron stars.  We  believe that it is possible 
to carry out the same procedure by decomposing the matter equation for
fermions, 
but it is beyond the scope of this work. 
Finally, for illustrative purposes, we present a simple case of study consisting in a couple of bosonic systems -plane-wave case and the spherically symmetric case-laying in flat space-time whose scalar equation is the Newtonian one plus a first order relativistic (post-Newtonian) correction which gives rise to non-static behavior of the potentials in the balance equation. 

\begin{acknowledgements}
This work was partially supported by CONACyT M\'exico under grants CB-2011 No. 166212, CB-2014-01 No. 240512, Project
No. 269652 and Fronteras Project 281;
Xiuhcoatl and Abacus clusters at Cinvestav, IPN; I0101/131/07 C-234/07 of the Instituto
Avanzado de Cosmolog\'ia (IAC) collaboration (http://www.iac.edu.mx/). We would like to 
acknowledge L.E. Padilla for useful comments and corrections in the elaboration of this
manuscript. T.B. and A.A. acknowledge financial support from CONACyT postdoctoral
fellowships.
\end{acknowledgements}

%%%%%%%%%%%%%%%%
% BIBLIOGRAPHY %
%%%%%%%%%%%%%%%% 

%\nocite{*}
\bibliographystyle{spphys}
\bibliography{GR_Boson_Gas}

\end{document}